\documentclass[conference]{IEEEtran}

\IEEEoverridecommandlockouts

\usepackage[cmex10]{amsmath}
\usepackage{amssymb}

\usepackage{algorithm, algpseudocode}
\usepackage{multirow}
\usepackage{graphicx}
\usepackage{xcolor}
\usepackage[capitalise]{cleveref}

\ifCLASSOPTIONcompsoc	\usepackage[caption=false,font=normalsize,labelfont=sf,textfont=sf]{subfig}
\else \usepackage[caption=false,font=footnotesize]{subfig}
\fi

\ifCLASSOPTIONcompsoc \usepackage[nocompress]{cite}
\else                 \usepackage{cite}
\fi

\begin{document}

\title{Context-aware Rate Adaptation for Predictive Flying Networks using Contextual Bandits\vspace{-0.3cm}
}

\author{

    \IEEEauthorblockN{
        Ruben Queiros\IEEEauthorrefmark{1},
        Megumi Kaneko\IEEEauthorrefmark{2},
        Helder Fontes\IEEEauthorrefmark{1},
        Rui Campos\IEEEauthorrefmark{1}
    }
    \IEEEauthorblockA{
        \IEEEauthorrefmark{1}INESC TEC, Faculty of Engineering, University of Porto, Porto Portugal 
         \\
        \IEEEauthorrefmark{2}National Institute of Informatics and Dept. of Computer Science, The University of Tokyo, Tokyo Japan, \\ \{ruben.m.queiros, helder.m.fontes, rui.l.campos\}@inesctec.pt, megkaneko@nii.ac.jp
    }   
    
    \thanks{This work is financed by National Funds through the Portuguese funding agency, FCT - Fundação para a Ciência e Tecnologia, under the PhD grant 2022.10093.BD, by the NII MoU grant between NII and INESC TEC, and by the Grant-in-Aid for Scientific Research no. 22KK0156 from the Ministry of Education, Science, Sports, and Culture of Japan.}
    \vspace{-1.2cm}

}

\maketitle

\begin{abstract}

The increasing complexity of wireless technologies, such as Wi-Fi, presents significant challenges for Rate Adaptation (RA) due to the large configuration space of transmission parameters. While extensive research has been conducted on RA for low-mobility networks, existing solutions fail to adapt in flying networks, where high mobility and dynamic wireless conditions introduce additional uncertainty. 

We propose Linear Upper Confidence Bound for RA (LinRA), a novel Contextual Bandit-based approach that leverages real-time link context to optimize transmission rates. Designed for predictive flying networks, where future trajectories are known, LinRA proactively adapts to obstacles affecting channel quality. Simulation results demonstrate that LinRA converges $\mathbf{5.2\times}$ faster than state-of-the-art benchmarks and improves throughput by 80\% in Non Line-of-Sight (NLoS) conditions, matching the performance of ideal algorithms. With low time complexity, LinRA is a scalable and efficient RA solution for predictive flying networks.

\end{abstract}

\vspace{-0.3cm}

\section{Introduction}
\label{sec:introduction}
\vspace{-0.1cm}

Next-generation wireless networks, including Wi-Fi~7/Wi-Fi~8, and beyond 5G/6G, are designed to support high-throughput, low-latency applications in highly dynamic environments. These advancements enable new use cases, such as Flying Networks (FNs), which consist of Unmanned Aerial Vehicles (UAVs) with communication and sensing capabilities. FNs provide scalable, on-demand connectivity in scenarios such as disaster response and infrastructure failures. However, despite their high mobility, rapid deployment, and adaptability, FNs introduce significant challenges in ensuring reliable communication under dynamic wireless conditions.

\begin{figure}
    \centering
    \includegraphics[trim={0.6cm 0.6cm 0.1cm 0.1cm}, clip, width=.8\linewidth]{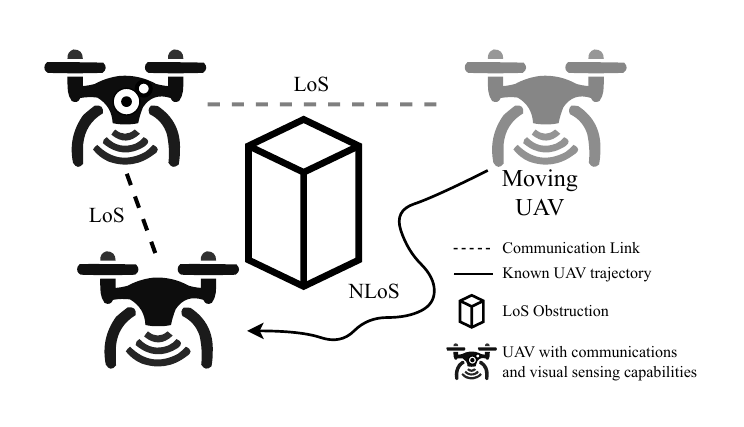}
    \vspace{-0.4 cm}
    \caption{\textbf{Predictive Flying Network example scenario.}}
    \label{fig:scenario}
    \vspace{-0.65 cm}

\end{figure}

A key challenge in FNs is Rate Adaptation (RA), which involves selecting the optimal Modulation and Coding Scheme (MCS) to maximize throughput while ensuring link reliability. RA has been extensively studied in IEEE 802.11 networks~\cite{mac80211survey}. Still, next generation networks introduce a significantly larger transmission configuration space, with multiple modulation schemes, wider bandwidths, and diverse spatial stream options, making RA decision-making increasingly challenging. Traditional RA algorithms often assume gradual channel variations and rely on static or heuristic-based approaches, making them unsuitable for highly dynamic mobile scenarios such as FNs. Yet, in predictive FNs, as illustrated in Fig.~\ref{fig:scenario}, UAV trajectories are known a priori, and onboard visual sensing (e.g., cameras) enables proactive obstacle detection. This contextual information can be used to optimize RA, improving spectrum utilization and ensuring robust link performance. Although prior research has explored UAV trajectory optimization and resource allocation~\cite{facom-survey}, context-aware RA in predictive FNs remains an open challenge.

Recent research has applied stochastic optimization in Multi-Armed Bandit (MAB) frameworks~\cite{combes2018optimal} to enhance RA in dynamic wireless conditions. Similarly,~\cite{tong2023rate} introduces Correlated Thompson Sampling (TS) and Correlated Kullback-Leibler Upper Confidence Bound (KL-UCB) RA algorithms. However, these approaches focus on terrestrial networks and assume gradual fading, making them less effective in FNs, where rapid link disruptions may occur.

Context-aware RA solutions for FNs have been proposed in~\cite{he2019state} and~\cite{xiao2021sensor}, using UAV motion data to improve RA decisions. \cite{he2019state} employs a deep learning-based prediction model for UAV motion but relies on offline data processing, while~\cite{xiao2021sensor} uses polynomial regression to estimate channel variations based on velocity and position. Joint optimization approaches, such as the scheduling framework in~\cite{wu2023joint}, integrate RA with virtual queues and UCB-based rate selection. However, these methods either depend on pre-collected data traces or lack the real-time adaptability needed for dynamic scenarios like FNs.

This paper proposes LinRA, a Linear UCB-based RA algorithm that formulates RA in predictive FNs as a Contextual Bandit (CB) problem. LinRA leverages contextual features, such as UAV trajectories and obstacle detection, to dynamically adapt the transmission rate in real-time. By employing LinUCB's exploration-exploitation mechanism, LinRA learns an optimal RA policy that balances throughput maximization and connection reliability under uncertain, time-varying wireless channel conditions.

The contributions of this paper are three-fold:
\begin{enumerate}
    \item{Context-aware RA formulation:} We introduce a novel RA problem where visual context information from UAV nodes enhances real-time wireless channel estimation;
    
    \item{Proposed Solution:} We develop LinRA, a LinUCB-based RA algorithm that dynamically selects the optimal transmission rate using contextual information while balancing exploration and exploitation;

    \item{Performance Analysis:} We provide numerical results demonstrating that LinRA outperforms State-of-the-Art (SotA) benchmarks in terms of throughput and convergence time.
    \vspace{-0.2cm}
    \end{enumerate}

\section{System Model} 
\vspace{-0.1cm}

We discretize time $T\in\mathbb{R}$ into $t$ time steps, where $t\in \mathbb{N}$. At each time step, the rate adaptation algorithm selects a MCS index from $I$ available transmission options. We denote this selection as $i_t \in \{1,...,I\}$, representing the chosen action for frame transmission at time step $t$, with a transmission duration $\tau_t$, which depends on the selected MCS and frame size. The decision-making process is guided by an \emph{N}-dimensional wireless environment context vector, denoted as $\mathbf{x}_t\in\mathbb{R}^N$. Finally, the transmission outcome $y_t\in \{0,1\}$ is a binary variable indicating success (1) or failure (0), which depends on both the selected MCS index $i_t$ and the context vector $\mathbf{x}_t$.

\vspace{-0.1cm}
\subsection{Channel Model}
\vspace{-0.1cm}

 In FNs, the wireless channel is predominantly characterized by a strong, obstacle-free Line-of-Sight (LoS) component~\cite{channel-uav-survey}. Therefore, we model the large-scale fading effects, specifically the path loss attenuation during time step $t$, using the Free Space Path Loss (FSPL) equation:\vspace{-0.1cm}
\begin{equation}
    L_{t}^{\mathrm{FSPL}}|_{\mathrm{dB}}(d_t)=20\log_{10}\left(\frac{\lambda}{4\pi d_t}\right),
    \label{eq:fspl}
    \vspace{-0.1cm}
\end{equation}
where $d_t$ represents the Euclidean link distance between the transmitter and the receiver, and $\lambda$ is the signal wavelength. Given that UAVs follow predefined mission trajectories, the link distance $d_t$ over time $T$ is assumed to be known as in~\cite{bowenli}.

To model shadowing effects caused by obstacles blocking the LoS, we adopt an empirical model inspired by~\cite{obstacle-model}. Instead of a stochastic approach, we use deterministic, measurement-based values for the additional signal attenuation $L^{\mathrm{obs}}_t$, modeled as a uniformly distributed random variable: $L^{\mathrm{obs}}_t|_{\mathrm{dB}}\sim U(o_\mathrm{min},o_\mathrm{max})$. We assume a blockage event model with two LoS periods and one Non-LoS (NLoS) period, as illustrated in Fig.~\ref{fig:scenario}, where the NLoS duration is randomly distributed but constrained within a minimum duration $\Delta t_\mathrm{min}$ and maximum bound $T$. During LoS periods, obstacle-induced attenuation is negligible, i.e., $L^{\mathrm{obs}}_t|_{\mathrm{dB}}=0$. For small-scale effects, we model multipath signal components using the Rician Fading distribution, characterized by the probability density function:
\begin{equation}
    p(x;\nu,\sigma)=\frac{x}{\sigma^2}\exp\left(\frac{-(x^2+\nu^2)}{2\sigma^2}\right)I_0\left(\frac{x\nu}{\sigma^2}\right), x\geq0,
    \label{eq:rician}
\end{equation}
where $I_0$ is the modified Bessel function of the zeroth order, and the parameters $\nu^2 = \kappa/(\kappa+1)$ and $2\sigma^2 = 1/(\kappa+1)$ define the fading characteristics based on the Rician $\kappa$-factor, which represents the ratio of LoS power to multipath power. The small-scale fading effects for each time step $t$ are denoted as $|h_t|^2 \sim \text{Rician}(\kappa)$, sampled from a Rician distribution.

We assume a block fading channel, where the channel state remains constant throughout the frame transmission duration $\tau_t$. Under these conditions, the received Signal-to-Noise Ratio (SNR) for a transmitted frame at time step $t$ is given by:\vspace{-0.1cm}
\begin{equation}
    \Gamma_t=\frac{P^{T} L_t |h_t|^2}{N_0 B} ,
    \label{eq:SNR}
    \vspace{-0.1cm}
\end{equation}
where $P^T$ is the fixed transmit power, $L_t = \frac{L_t^{\mathrm{FSPL}}}{L_t^{\mathrm{obs}}}$ represents the total path loss, incorporating both FSPL $L_t^{\mathrm{FSPL}}$ and obstacle-induced $L_t^{\mathrm{obs}}$, $|h_t|^2$ captures the small-scale fading effects, $N_0$ is the power spectral density of Additive White Gaussian Noise (AWGN), and $B$ denotes the channel bandwidth, assumed to be managed by a SotA resource allocation algorithm such as the one proposed in~\cite{SAFnet}.

\vspace{-0.1cm}
\subsection{Rate Model}
\vspace{-0.1cm}

Data transmission is performed using Orthogonal Frequency Division Multiplexing (OFDM), the transmission method adopted in IEEE 802.11 networks. Let $(B,i)$ represent a supported link configuration, where $B$ is the channel bandwidth, and $i$ denotes the selected MCS index, with $i \in \{1,...,I\}$. The achievable Physical Data Rate $r$ (in bit/s) is expressed as:\vspace{-0.1cm}
\begin{equation}
    r(B,i)=\frac{N_{\mathrm{SS}} N_{\mathrm{DS}}(B) C_\mathrm{R}(i) C_\mathrm{B}(i)}{T_\mathrm{S}+T_{\mathrm{GI}}},
    \label{eq:phy-rate}
    \vspace{-0.1cm}
\end{equation}
where $N_{\mathrm{SS}}$ is the number of spatial streams, $N_{\mathrm{DS}}$ is the number of data subcarriers, which depends on the assigned channel bandwidth $B$, $C_\mathrm{R}(i)$ is the coding rate associated with the MCS index $i$, $C_\mathrm{B}(i)$ represents the number of coded bits per subcarrier per spatial stream, $T_\mathrm{S}$ is the OFDM symbol duration, and $T_\mathrm{GI}$ the guard interval duration. Without loss of generality, we consider a Single Input Single Output (SISO) configuration with a long Guard Interval and a fixed channel bandwidth. Thus, for simplicity, hereafter we denote the achievable rate as $r(i_t)$. The throughput in bit/s of a frame transmission is defined as:\vspace{-0.1cm}
\begin{equation}
    \eta(r,S,\Gamma)=\frac{\theta(r, \Gamma) S}{\tau (r,S)},
    \label{eq:thp}
    \vspace{-0.1cm}
\end{equation} 
where $S$ is the frame size in bits, $\tau (r,S)$ is the frame transmission duration, and $\theta(r, \Gamma)$ denotes the probability of successfully receiving a frame using rate $r$, with $\Gamma$ unknown \textit{a priori}.\vspace{-0.1cm}

\vspace{-0.2cm}
\section{Problem Formulation}
\vspace{-0.1cm}

Our objective is to design an algorithm that maximizes the expected link throughput by leveraging both wireless environment context and past transmission outcomes. A common approach to formulating decision-making problems in adaptive systems is through the concept of regret, which in our case quantifies the performance loss incurred due to suboptimal rate selection. The regret is mathematically formulated as:
\begin{align}
    \label{eq:prob}
        & \min_{i}\sum_{t=0}^{F_\mathrm{max}} [r(i^\star_t)y_t^\star-r(i_t)y_t], \\
        \label{eq:c-tx_time}
        & \text{s.t.}~ \sum_{t=0}^{F_\mathrm{max}} \tau_t(r(i_t),S) \leq T \\
        &\forall i\in \{1,...,I\}, \forall y \in 
        \{0,1\}, \forall t\in \mathbb{N}
\end{align}
where $i^\star_t$ denotes the optimal MCS index at time step $t$, $y^\star_t$ is the corresponding transmission success indicator, $i_t$ is the MCS index selected by the RA algorithm, and $y_t$ represents the actual transmission success outcome at time step $t$. Finally, Eq.~(\ref{eq:c-tx_time}) constraints the total number of transmitted frames $F_\mathrm{max}$ to be transmitted within $T$.  

\vspace{-0.1cm}
\section{Proposed Solution: LinUCB for Rate Adaptation (LinRA)}
\vspace{-0.1cm}

The LinUCB algorithm~\cite{linucb} extends the Upper Confidence Bound algorithm, designed for contextual bandit problems. Unlike standard multi-armed bandit approaches, LinUCB incorporates contextual information to make more informed decisions. The underlying idea is that the expected reward for each action (arm) is modeled as a linear function of contextual features, which the algorithm learns over time. This enables LinUCB to adapt to environmental changes while balancing exploration (selecting less-tested arms) and exploitation (selecting arms with high expected rewards). 

We propose LinRA, a LinUCB-based rate adaptation algorithm that optimizes MCS selection to maximize link throughput. The algorithm leverages link-specific features, such as estimated link distance and obstacle detection, to predict the expected throughput reward, defined as:\vspace{-0.1cm}
\begin{equation}
R_t= \frac{r(i_t) y_t}{r(I)},
\label{eq:rw}
\vspace{-0.1cm}
\end{equation}
where $R_t$ is the reward at time step $t$, $r(i_t)$ is the data rate associated with the selected MCS index $i_t$, $y_t$ is a binary indicator of transmission success, and $r(I)$ is the maximum achievable rate used to normalize the reward value. The \textbf{context vector} at time step $t$ is given by $\mathbf{x}_t^T = \begin{bmatrix} d_t & \mathrm{F}^\Omega_t  \end{bmatrix}$, where $d_t$ is the link distance between nodes at time step $t$ and $\mathrm{F}^\Omega_t$ is a boolean flag indicating the presence of an obstacle. Since $F_t^\Omega$ does not quantify the magnitude of the impact on the link, it serves purely as a binary indicator of obstacle. These contextual features are obtained from mission knowledge~\cite{bowenli} and on-board cameras~\cite{respondrone}. Mission knowledge provides UAV node positions, allowing link distance estimation; cameras allow obstacle detection, enabling the system to anticipate LoS and NLoS transitions.

\begin{algorithm}
\caption{LinRA: LinUCB for Rate Adaptation}
\begin{algorithmic}[1]

\State \textbf{Initialize:} $\mathbf{A}_i = \mathbf{I}_{N}, \mathbf{b}_i=\mathbf{0}_{N},\alpha=1, d_\mathrm{max}=0$

\For{$t=1,2,...,F_\mathrm{max}$}
\State $d_\mathrm{max} \leftarrow \max(d_t, d_\mathrm{max}) $
\State $\mathbf{x}_t \leftarrow \begin{bmatrix} d_t/d_\mathrm{max} & \mathrm{F}^\Omega_t  \end{bmatrix}$

\If{$\mathrm{F}^\Omega_t \neq \mathrm{F}^\Omega_{t-1}$} 
$\alpha \leftarrow 1$
\EndIf \text{ // Obstacle status changed}

\For{$i=1,...,I$}

\State $p_{t,i} \leftarrow (\mathbf{A}_i^{-1} \cdot \mathbf{b}_i)^\top \cdot \mathbf{x}_t + \alpha\sqrt{\mathbf{x}_t^\top\cdot\mathbf{A}_i^{-1}\cdot\mathbf{x}_t}  $

\EndFor

\State $i_t^\star \leftarrow \arg\max_i(p_{t,i})$ \text{// Select highest UCB MCS}

\State $R_t \leftarrow r(i^\star_t)y^\star_t/ r(I)$\textbf{ // After transmission}
\State $\mathbf{A}_{i^\star} \leftarrow \mathbf{A}_{i^\star} + \mathbf{x}_t \cdot \mathbf{x}_t^\top$
\State $\mathbf{b}_{i^\star} \leftarrow \mathbf{b}_{i^\star} + R_t \cdot \mathbf{x}_t$
\State $\alpha \leftarrow \alpha \times \epsilon$ \text{// Decay exploration factor}

\EndFor

\end{algorithmic}

\label{alg:linra}

\end{algorithm}

The LinRA algorithm, detailed in Algorithm~\ref{alg:linra}, works as follows. It starts by initializing arm-specific parameters, including the feature covariance matrix \(\mathbf{A}_i\) and reward vector \(\mathbf{b}_i\), which track the relationship between context features and rewards for each MCS index (Line 1). At each time step \(t\), the algorithm constructs the context vector \(\mathbf{x}_t\) by computing the current link distance $d_t$, dynamically normalizing it based on the maximum observed distance $d_\mathrm{max}$, and retrieving the obstacle presence flag $\mathrm{F}^\Omega_t$ (Line 3). If the obstacle flag $F_t^\Omega$ changes state, the exploration parameter $\alpha$ is reset to its initial value $\alpha=1$ (Line 5). For each MCS index (arm) $i$, the algorithm computes the UCB score \(p_{t,i}\) (Line 8), which combines the estimated model parameters and an exploration term scaled by \(\alpha\). The MCS index with the highest UCB score is then selected for transmission (Line 10). After transmission, the observed reward is  calculated, based on Eq.~(\ref{eq:rw}) (Line 12), and used to update the parameters related to the selected MCS index (Lines 13-14). The exploration parameter \(\alpha\) decays over time (Line 15), gradually shifting the strategy from exploration to exploitation. This approach enables LinRA to dynamically select the optimal transmission rate in real-time while continuously improving its decision-making based on observed context and outcomes.

\begin{table}
\vspace{-0.35cm}
    \centering
    \caption{Simulation Parameters}
    \vspace{-0.25 cm}

    \begin{tabular}{lc|l}
         Description & Symbol & Value\\
         \hline
         Simulation duration (s) & $T$ & 30 \\

         Transmit Power (dBm) & $P^T$ & 20 \\
         Wavelength (mm) & $\lambda$ & 125 \\
         Channel Bandwidth (MHz) & $B$ & 20 \\
         Power Spectral Density (dBm/Hz) & $N_0$ & -174 \\
         Rician K-Factor (dB) & $\kappa$ & 13 \\ %
         Maximum Physical Data Rate (Mbit/s) & $r(I)$ & 65 \\
         Frame Size (bytes)\cite{ns3-error-model} & $S$ & 1458 \\
         NLoS obstacle attenuation (dB)\cite{obstacle-values}\cite{obstacle-model} & $L^{\mathrm{obs}}|_{\mathrm{dB}}$ & $U(10,15)$  \\ 
         NLoS period min duration (s) & $\Delta t_{\mathrm{min}}$ & 2 \\
    \end{tabular}
    \vspace{-0.6 cm}
    \label{tab:simulation-param}
\end{table}

\vspace{-0.1cm}
\section{Performance Evaluation}
\vspace{-0.1cm}

\vspace{-0.1cm}
\subsection{Rate Adaptation Benchmarks}
\label{sec:refmethods}
\vspace{-0.1cm}

LinRA is evaluated against four representative RA benchmarks: Thompson Sampling (TS), a Random decision algorithm (baseline), and two idealized algorithms -- Oracle and Semi-Oracle. The \textbf{TS}-based RA algorithm, originally proposed in~\cite{ts}, applies a Bayesian approach to dynamically adapt the MCS selection. At each time step $t$, TS samples success probabilities $X_{t,i}$ from a Beta distribution, parameterized by the success count $a_i$ and failure count $b_i$ for each MCS index $i$. The expected reward for each MCS is computed as $X_{t,i}\sim\mathrm{Beta}(a_i,b_i)$ and $i_t^\star = \arg\max_i(r(i_t)X_{t,i})$, where $r(i_t)$ is the corresponding physical data rate, as defined in Eq.~(\ref{eq:phy-rate}). After each transmission, the observed outcome $y_t$ is used to update the Beta distribution parameters using exponential smoothing, ensuring adaptability to changing conditions. The update rule is given by:
\begin{equation}
a_i(t) = \left\{ 
\begin{array}{ll}
a_i(t - \Delta t)e^{-\frac{\Delta t}{w}}+1 &, y_t=1 \\
a_i(t - \Delta t)e^{-\frac{\Delta t}{w}} &, y_t=0 
\end{array} 
\right.
\label{betaparams}
\end{equation}
where $w$ is the exponential window controlling the decay rate, balancing recent observations with historical trends. A similar update rule applies to $b_i$ when a transmission failure occurs.

The \textbf{Random} algorithm selects an MCS index $i_t$ at random from a discrete uniform distribution $U\{1,I\}$, serving as a baseline to establish the minimum performance that any effective RA algorithm should surpass.
The \textbf{Oracle} algorithm assumes perfect knowledge of future transmission channel conditions, including small-scale fading modeled as a Rician fading distribution and pre-computed error probability tables~\cite{ns3-error-model}. As a result, Oracle achieves a 100\% frame success ratio by always selecting the optimal MCS.
The \textbf{Semi-Oracle} algorithm, while similar to Oracle, can only predict future channel conditions based on large-scale effects such as FSPL and obstacles but does not account for instantaneous small-scale fading variations. Despite being a weaker variant of Oracle, Semi-Oracle retains the error probability knowledge.

\subsection{Simulation Setup}
\vspace{-0.1cm}

The flying network operates within a three-dimensional coverage area of $1000\times 1000\times20$ meters. We evaluate throughput in terms of successful frame reception, as defined in Eq.~(\ref{eq:thp}), rather than at symbol level, using frame error ratio tables derived from simulation results~\cite{ns3-error-model}. The custom-tailored simulator
used\footnote{https://gitlab.inesctec.pt/pub/ctm-win/linra\vspace{-0.1cm}} is publicly available. For RA algorithm simulations, key parameters were empirically set, including LinRA's decay rate $\epsilon=10^{-3}$ and TS's exponential window $w=1$. Additional fixed parameters are listed in Table~\ref{tab:simulation-param}. 
A unique random seed is used for each simulation, affecting the initial positions and trajectories of network nodes. Despite this randomness, all trajectories are constrained within the defined coverage zone.

We evaluate convergence time and throughput. \textbf{Convergence time} is defined only for learning-based RA algorithms, such as LinRA and TS, as the time elapsed between the channel change event and the start of the first 1-second window where throughput consistently remains within 95\% of the reference algorithm. Non-learning RA benchmarks, such as Random, Semi-Oracle, and Oracle, do not have a convergence time, since they do not adapt their decisions based on past observations. \textbf{Throughput} is evaluated at three key moments:

\begin{enumerate}
    \item \textbf{Reaction throughput} is measured during the first 1-second window of either the NLoS period or the second LoS period. 
    \item \textbf{Convergence throughput} is computed over the time interval starting from an LoS-to-NLoS or LoS-to-NLoS transition until the slowest learning-based RA algorithm converges. 
    \item \textbf{Stability throughput} is measured during the last 1-second window of the NLoS period or second LoS period, representing the final system performance.
\end{enumerate}

\vspace{-0.1cm}
\subsection{Simulation Results}
\vspace{-0.1cm}
\begin{figure}
    \centering
    \includegraphics[trim={0 0cm 0 1cm}, clip, width=.85\linewidth]{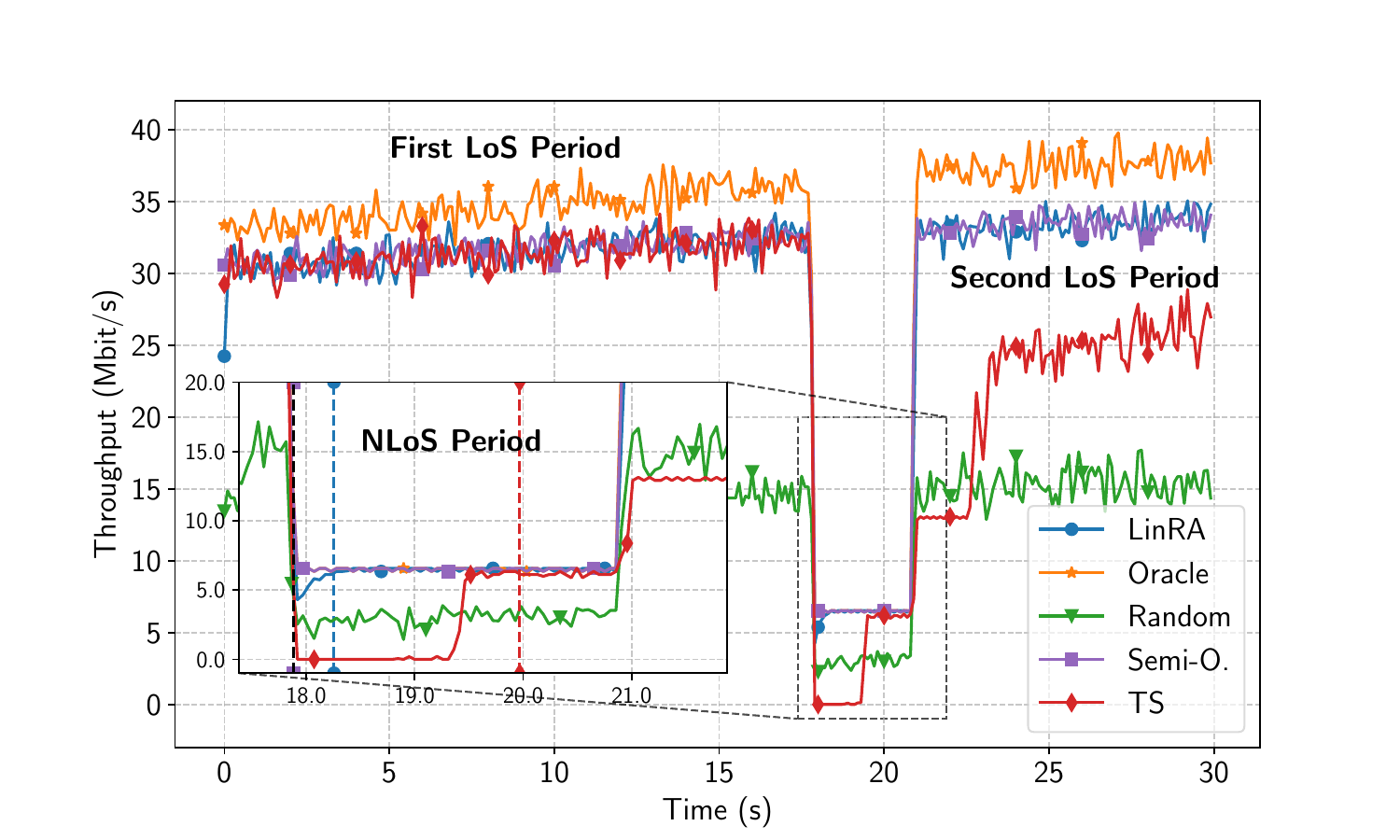}
    \vspace{-0.45 cm}
    \caption{Comparison of RA algorithms throughput for a specific random seed.}
    \label{fig:throughputovertime}
    \vspace{-0.6 cm}

\end{figure}

\begin{figure*}
  \centering
    \subfloat[
Average reaction throughput.\label{fig:NLOS_REACT}]{
      \includegraphics[trim={0 0cm 0 1cm}, clip, width=.32\textwidth]{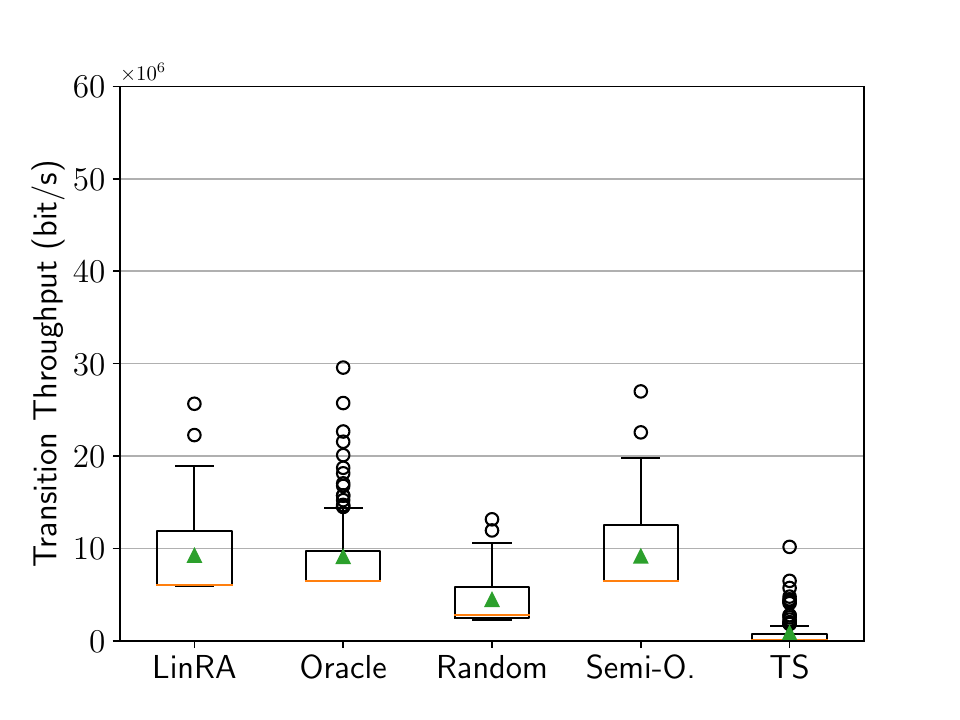}   
  }
  \hspace{-0.8 cm}
  \subfloat[Average convergence throughput.\label{fig:LOS_CONV}]{
      \includegraphics[trim={0 0cm 0 1cm}, clip, width=.32\textwidth]{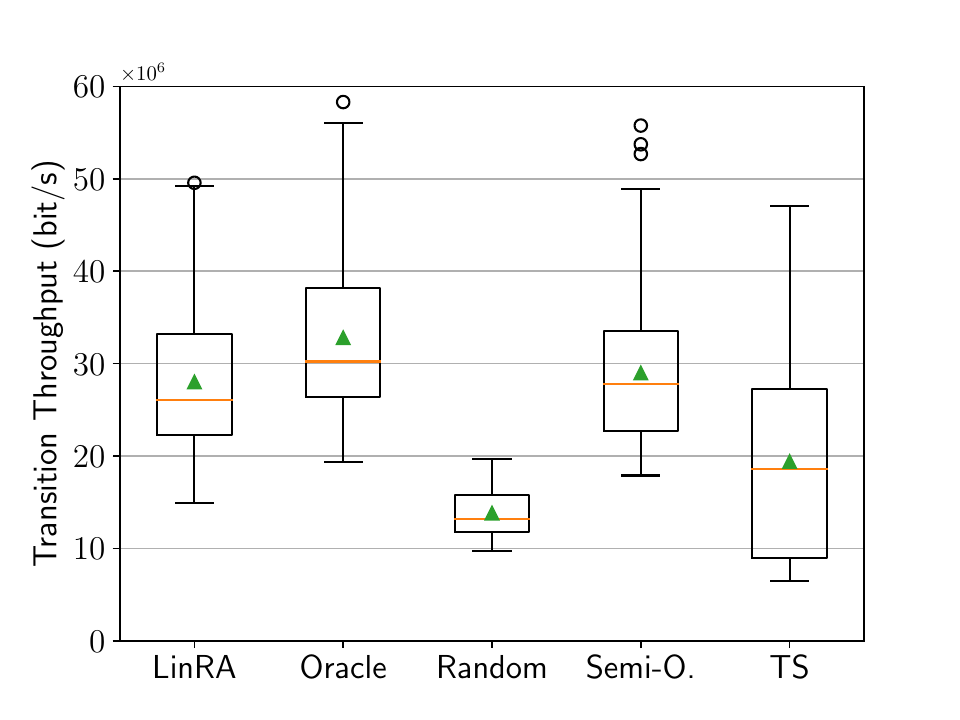}
  }
  \hspace{-0.8 cm}
  \subfloat[Average stability throughput.\label{fig:LOS_STAB}]{
      \includegraphics[trim={0 0cm 0 1cm}, clip, width=.32\textwidth]{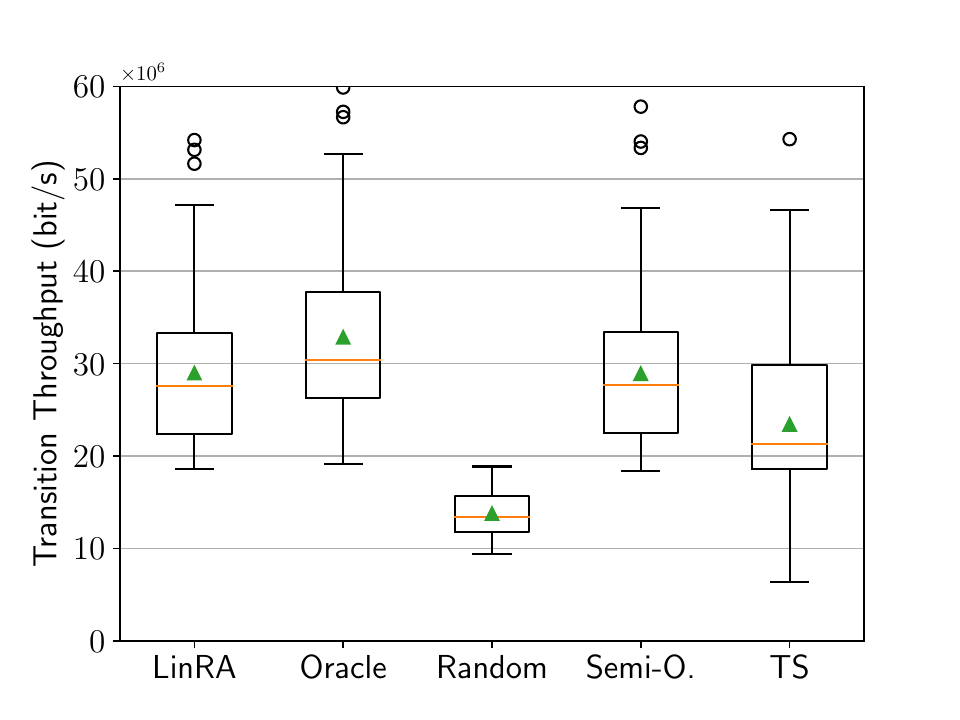}
  }
  \caption{Average throughput of the RA algorithms over 100 simulation using different seeds, focusing on the three transition moments.}
  \vspace{-0.6 cm}
  \label{fig:ext}
\end{figure*}

We start with a scenario analysis using a single random seed. Fig.~\ref{fig:throughputovertime} shows the throughput over time, with a NLoS period lasting approximately 3 seconds. During the first LoS period, throughput differences are negligible, except for the Random algorithm, which performs significantly worse. Since most existing works assume smooth channel variations, they overlook the importance of reaction to sudden channel changes, which is the focus of LinRA (as discussed in Sec.~\ref{sec:introduction}). Therefore, our analysis is focused on the NLoS and second LoS periods, where the most significant channel condition changes occur. 

Fig.~\ref{fig:throughputovertime} provides a detailed view of the LoS-to-NLoS transition, where the NLoS period ocurrs for $ t \in [17.9, 20.9]$ seconds, with dashed lines indicating the convergence time of the RA algorithms. LinRA achieves a convergence time of $375$~ms while TS takes $5.6\times$ longer. In terms of reaction throughput, LinRA achieves 6~Mbit/s, while TS remains close to 0~Mbit/s due to its convergence time exceeding 1 second in this scenario. For the convergence throughput, measured for $t \in [17.9, 20]$ seconds, LinRA achieves 6.3~Mbit/s, significantly outperforming TS, which reaches only 1.5~Mbit/s. Finally, the stability throughput of TS recovers to 6.2~Mbit/s, while LinRA remains nearly identical to Oracle at 6.5~Mbit/s. A similar trend is observed during the recovery transition, where TS fails to converge, and LinRA outperforms it.

We now present results aggregated over 100 random seeds, summarized in Fig.~\ref{fig:ext} and Table~\ref{tab:ext-results}. LinRA achieves  convergence in 99\% of the NLoS periods and 75\% of LoS periods. On the other hand, TS only converges in 64\% and 18\% of cases, respectively. Concerning average convergence time, LinRA recovers $5.2\times$ faster than TS in NLoS periods and $2.1\times$ faster in LoS periods, with average recovery times of $335$~ms and $959$~ms, respectively. Fig.~\ref{fig:ext} presents the average throughput of the RA algorithms at different transition moments. The boxplot shows the first and third quartiles, with an orange line and green triangle indicating the median and mean, respectively. The whiskers extend by $1.5\times$ the inter-quartile range, with outliers marked as points beyond the whiskers. Fig.~\ref{fig:ext}\subref{fig:NLOS_REACT} highlights the worst reaction-phase throughput performance by TS during NLoS, where its convergence time typically exceeds 1 second, leading to near zero transition throughput. LinRA and Semi-Oracle outperform Oracle in certain cases due to Oracle's design: perfect channel knowledge occasionally leads to a lower MCS selection, which extends transmission times and slightly reduces throughput, despite a $100\%$ frame success ratio, as derived from Eq.~(\ref{eq:thp}). This effect does not occur in LoS periods, where higher average MCS indexes lead to increased throughput, as seen in Figs. \ref{fig:ext}\subref{fig:LOS_CONV} and \ref{fig:ext}\subref{fig:LOS_STAB}. In LoS conditions, Oracle remains the best RA algorithm, while LinRA consistently outperforms TS across both transition moments.

\begin{table}
\centering
\caption{Mean Throughput normalized to Oracle's throughput.}
    \vspace{-0.2 cm}

\begin{tabular}{c|ccc|cccll}

\multirow{2}{*}{RA Alg.} & \multicolumn{3}{c|}{NLoS Period} & \multicolumn{3}{c}{Second LoS Period} &  &  \\ \cline{2-7}
                              & React.          & Stab.         & Conv.         & React.          & Stab.         & Conv.         &  &  \\ \cline{1-7}
Random                        & 0.48            & 0.48          & 0.48          & 0.43            & 0.43          & 0.43          &  &  \\ \cline{1-7}
Semi-Oracle                   & 1.01            & 1.00          & 1.00          & 0.88            & 0.88          & 0.89          &  &  \\ \cline{1-7}
TS                            & 0.06            & 1.00          & \textbf{0.20}          & 0.54            & 0.70          & \textbf{0.56}          &  &  \\ \cline{1-7}
LinRA                         & 1.02            & 1.10          & \textbf{1.00}          & 0.87            & 0.88          & \textbf{0.85}          &  & \end{tabular}
\label{tab:ext-results}
\vspace{-0.7 cm}

\end{table}

Table~\ref{tab:ext-results} presents the normalized throughput, expressed as the ratio of a given RA algorithm's throughput to Oracle's throughput. LinRA remains the most reliable method meeting Oracle's performance while being implementable and consistently outperforming TS in both NLoS and LoS periods. In particular, LinRA achieves $5\times$ higher mean convergence throughput in NLoS (1.00 vs. 0.20) and a 52\% improvement in the second LoS (0.85 vs. 0.56) compared to TS.

\vspace{-0.1cm}
\subsection{Time Complexity Analysis}
\vspace{-0.1cm}
The time complexity of TS is $\mathcal{O}(I)$ during rate selection, meaning it scales linearly with the number of MCS indexes. Conversely, LinRA has a time complexity of $\mathcal{O}(N^3I)$, due to the matrix inversions and computations associated with $\mathbf{A}_i$. While TS is computationally more efficient, LinRA enables more robust decision-making, particularly in scenarios where the relationship between context and rewards is critical. LinRA is especially suitable when the context vector dimension $N$ is small enough to meet real-time computational constraints.

\vspace{-0.1cm}
\section{Conclusions and Future Work}
\vspace{-0.1cm}

Predictive FNs require novel RA solutions to handle dynamic wireless conditions effectively. This paper proposed LinRA, a LinUCB-based RA algorithm that leverages contextual information, including link distance estimation and obstacle detection, to optimize transmission rate selection. By proactively adapting to sudden channel changes, LinRA significantly reduces the convergence time, outperforming SotA benchmarks while being implementable and computationally feasible. Future work will assess the impact of delayed contextual data on LinRA's performance and validate the proposed solution experimentally.

\vspace{-0.4cm}
\bibliographystyle{IEEEtran}
\bibliography{refs}
\end{document}